\documentclass[twocolumn]{aastex63}

\usepackage{tikz, amsmath, txfonts, comment}
\hypersetup{linkcolor=blue, citecolor=blue}
\usetikzlibrary{decorations.pathreplacing}
\usetikzlibrary{arrows.meta}

\received{date}
\revised{date}
\accepted{date}

\submitjournal{ApJ}

\shorttitle{The Interstellar Carbon K-Edge}
\shortauthors{Staunton and Paerels}

\begin{document}

\title{X-ray Spectroscopy of Interstellar Carbon: \\ Evidence for Scattering by Carbon-Bearing Material in the Spectrum of 1ES 1553+113}

\author{John Staunton}
\affiliation{Columbia Astrophysics Laboratory, Columbia University, 538 W. 120th St., New York, NY, 10027, USA.}

\author{Frits Paerels}
\affiliation{Columbia Astrophysics Laboratory, Columbia University, 538 W. 120th St., New York, NY, 10027, USA.}

\begin{abstract}
Molecules and particles make up $\sim 40 - 70\%$ of carbon in the interstellar medium, yet the exact chemical structure of these constituents remains unknown. We present carbon K-shell absorption spectroscopy of the Galactic Interstellar Medium obtained with the Low Energy Transmission Grating Spectrometer on the {\it Chandra} Observatory, that directly addresses this question. We probe several lines of sight, using bright AGN as backlighters. We make our measurements differentially with respect to the bright source
Mrk 421, in order to take the significant carbon K absorption in the instrument into account. In the spectrum of the blazar 1ES 1553+113 we find evidence for a novel feature: strong extinction on the low-energy side of the neutral C $1s-2p$ resonance, which is indicative of
scattering by graphite particles. We find evidence for characteristic particle radii of order $0.1-0.15$ $\mu$m.
If this explanation for the feature is correct, limits on the mass of the available carbon along the line of sight may imply that the grains are partially aligned, and the X-rays from the source may have intrinsic polarization.
\end{abstract}



\section{Introduction}
\label{S:1}

\subsection{Background}
\label{S:1.1}







The state of carbon in the interstellar medium (ISM) has been the subject of intense study \citep[e.g.][]{16, 28, 41, 4, 5, 6, 34, 36}. It has been determined that $\sim 40 - 70\%$ of the carbon in the ISM is in molecular or particulate form \citep{5}. The evidence for the existence of molecular carbon in the ISM is supported by the extinction bump at 2175 \AA. However, it is not obvious what specific molecular form the carbon takes as both Polycyclic Aromatic Hydrocarbons (PAHs) and graphite are viable explanations for this particular observation \citep{7, 38}.  

Innershell X-ray spectroscopy holds considerable promise in this respect. The X-ray spectrum, if observed at sufficient resolution, is sensitive to ionization as well as chemical bond effects, while covering all relevant species in a relatively compact wavelength band that can be observed all at once. In addition, X-ray absorption spectroscopy is not limited to specific environments ({\it e.g.} dense molecular clouds \citep{37}).

Carbon K-shell absorption is prominent in soft X-ray sources with moderate interstellar absorption, so that the ISM remains transparent down to photon energies below $\sim 280$ eV. 
Moreover, due to newfound interest in graphite and graphene in condensed matter physics, we now have access to better laboratory K-shell spectroscopy for materials that are often cited as candidates for compounds that make up the ISM \citep{5}.

The Low Energy Transmission Grating Spectrometer (LETGS) aboard the \textit{Chandra} X-ray Observatory has sufficient sensitivity and resolving power covering the C K spectrum \citep{35}. However, the LETGS has a UV/ion shield (UVIS) containing plastic, which in turn contains carbon. As a result, the main carbon absorption feature in astrophysical spectra is dominated by the instrument's response, as opposed to the interstellar carbon. Therefore, the first goal of this paper is to develop a model to accurately account for the instrument contribution to the observed spectrum around the carbon K edge.

In order to account for the instrument response around the carbon edge, we use an extremely bright AGN with known and very small Galactic foreground absorption, and remove all known contributions to the absorption and X-ray continuum shape, other than the UVIS filter transmission. We can then compare to other continuum sources which are known to have significant ISM absorption. The selection of our sample is discussed in \hyperref[S:1.2]{Section 1.2}, observations and data reduction in \hyperref[S:2]{Section 2}, and the empirical instrument response used for the remainder of this work in \hyperref[S:3]{Section 3}.

The second goal of this paper is to determine whether there are novel features in the ISM not previously studied by applying the empirical instrument response to an obscured source. In doing so, we report the first detection of $1s - 1\pi^*$ discrete absorption due to neutral interstellar C, and interpret the shape of the absorption feature in terms of scattering by small graphite spheres in \hyperref[S:4]{Section 4}.

We discuss our findings in \hyperref[S:5]{Section 5}. There, we examine the arguments that point to an interpretation in terms of scattering by graphite spheres, and examine possible systematics and non-astrophysical origins of the signal.

\subsection{X-ray Sample}
\label{S:1.2}






\begin{figure*}
    \centering
    \includegraphics[scale=0.36]{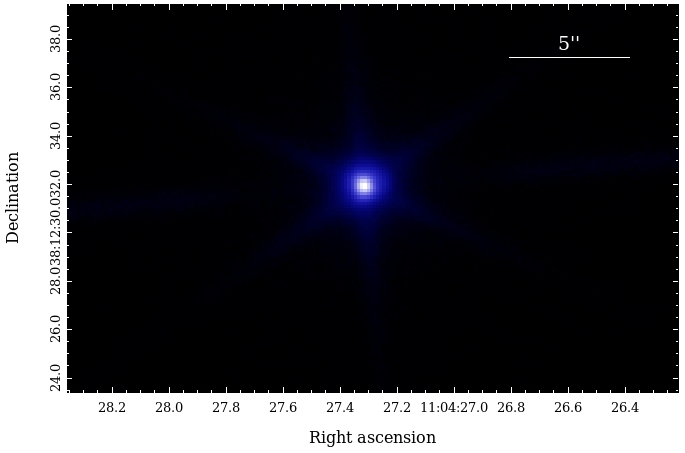}\includegraphics[scale=0.36]{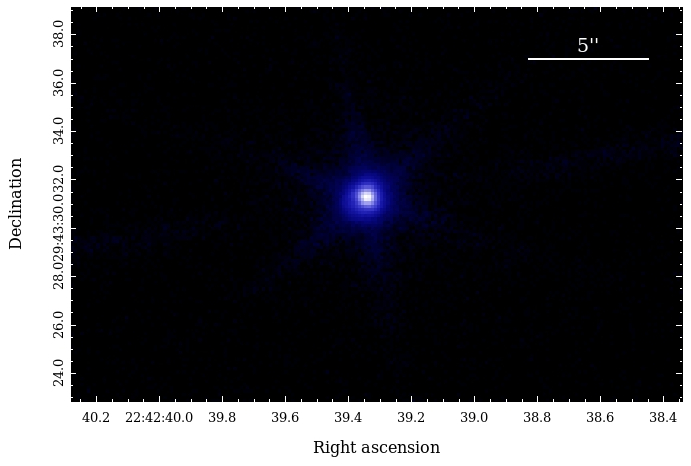}\\
    \includegraphics[scale=0.36]{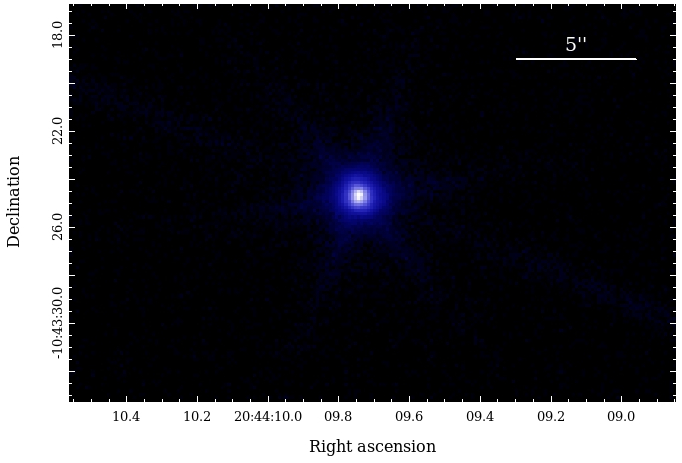}\includegraphics[scale=0.36]{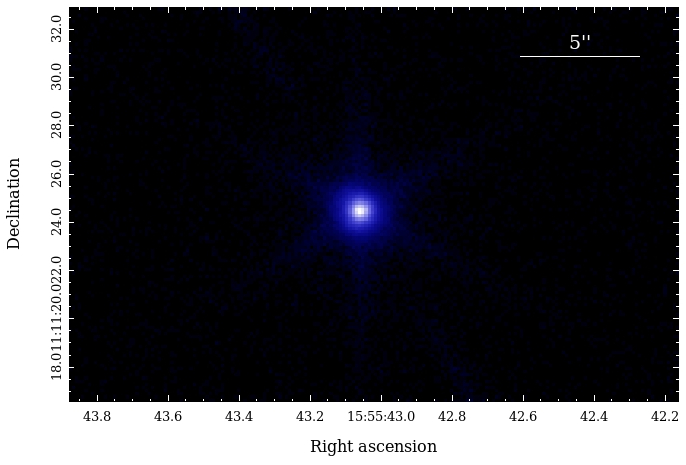}
    \caption{Objects of Study (0.2 - 8 keV). These are 0.2 - 8 keV images of Mrk 421 (top left), Ark 564 (top right), Mrk 509 (bottom left), and 1ES 1553 (bottom right). 
    The legend shows the length of 5 arcseconds. Colors, from black to blue to white, correspond to logarithmically scaled count number. The sources are all point-like and are located in empty fields.}
    \label{fig:1}
\end{figure*}

In order to carry out the procedure previously laid out, we require one high signal-to-noise source with low interstellar absorption observed with the LETGS to use as a proxy for an empirical instrument response. 
These requirements are satisfied by Mrk 421, a bright BL Lac object with a featureless, flat intrinsic X-ray continuum. There is a large amount of {\it Chandra} grating spectroscopy on the object; it has been used extensively to search for intergalactic absorption \citep{39}. The source has a low Galactic foreground neutral gas column density of $1.92 \times 10^{20}$ H atoms cm$^{-2}$ \citep{2}. The carbon K edge in the LETGS spectrum has been studied in detail by 
\citet{4}, who constructed a model for the UVIS filter transmission based on theoretical
model calculations for the X-ray absorption cross section of the neutral C atom. The spectrum of Mrk 421 is relatively hard, which means that we will have to pay detailed attention to the treatment of the higher-order contributions to the grating spectrum (\hyperref[S:3]{Section 3}). The top left panel \autoref{fig:1} shows the LETGS zeroth order spectral image of Mrk 421, from energies $0.2 - 8$ keV. As is apparent, there is no diffuse emission contaminating our spectrum and the source completely dominates the background.


The ideal candidate for our study is 
bright enough to have a high contrast absorption spectrum, yet has enough absorption so that it has an appreciable interstellar carbon K-edge in comparison to Mrk 421, meaning it has a column density above $\sim 3 \times 10^{20}$  cm$^{-2}$. We identified three potentially interesting objects due to their relatively high flux and column density: 1ES 1553+113, Mrk 509, and Ark 564. In addition to being bright, they were observed with long exposures, between 100 and 500 ks, making them ideal objects of study. All $0.2 - 8$ keV and $0.2 - 1.5$ keV images of these objects are presented in \hyperref[fig:1]{Figure 1}. 
The sources are point-like and there is no appreciable background contamination.

1ES 1553+113, henceforth referred to as 1ES 1553, is a blazar that emits strongly in both the gamma ray and X-ray regime \citep{18}. 1ES 1553 has a bright soft X-ray continuum and it has a column density appreciably higher than Mrk 421, at $N_{\rm H} = 3.65 \times 10^{20}$ H atoms cm$^{-2}$ \citep{2}, and it is therefore a good candidate for analysis of the carbon K-edge. As should be expected, however, it is not as bright as Mrk 421 and so sensitivity issues will arise as pointed out by \citet{36}.

Mrk 509 is a well-studied and extremely bright Seyfert galaxy \citep{42}. It is also extremely close with a redshift $z \sim 0.03$, compared to 1ES 1553 with a redshift of $z \sim 0.36$. In addition, it has a column density of $N_{\text{H}} = 4.17 \times 10^{20}$ cm$^{-2}$ \citep{2}, which is significantly higher than Mrk 421.

Ark 564 is a bright Seyfert 1 galaxy with a prominent soft X-ray excess that has been a prominent topic of study in recent years \citep{43}. It is even closer than the previous galaxies with $z = 0.025$ and has an appreciable column density of $N_{\text{H}} = 5.41 \times 10^{20}$ cm$^{-2}$. This column density is the highest of those previously mentioned, meaning it provides a high likelihood of displaying carbon features not present in the instrument itself.


\section{Observations and Data Reduction}
\label{S:2}

All data are based on observations made with the Low Energy Transmission Grating (LETG) and the HRC-S camera aboard the \textit{Chandra} X-ray observatory. This configuration is well suited for high resolution spectroscopy especially at low energies between $0.08$ and $1$ keV\footnote{For more info, visit http://chandra.harvard.edu/}. All data were obtained from the publicly available \textit{Chandra} data archive. We list the specific observation IDs and exposures in \hyperref[tab:1]{Table 1}. We used the \textit{Chandra} Interactive Analysis of Observations (CIAO) v4.9 package with calibration database 4.7.8 to reproduce event files and extract 8 positive and negative grating order spectra \citep{20}. Data used for calculating extinction models for graphite come from data presented in \citet{5}.

\begin{deluxetable}{ccccc}
\tablecaption{Observations\label{tab:1}}
\tablehead{
\colhead{ObsID} & \colhead{Galaxy} & \colhead{Date} & \colhead{Exposure (ks)} & \colhead{$N_H$ (cm$^{-2}$)$^{1}$}\\\vspace{-0.5cm}}
\decimals
\startdata
4149 & Mrk 421 & 2003-07-01 & 99.04 & $1.92 \times 10^{20}$ \\
11387 & Mrk 509 & 2009-12-10 & 131.46 & $4.17 \times 10^{20}$ \\
11388 & Mrk 509 & 2009-12-12 & 48.8 & \\
12915 & 1ES 1553 & 2011-05-04 & 166.31 & $3.65 \times 10^{20}$ \\
12916 & 1ES 1553 & 2011-06-18 & 153.94 & \\
12917 & 1ES 1553 & 2011-05-06 & 175.40 & \\
9151 & Ark 564 & 2008-04-21 & 99.96 & $5.41 \times 10^{20}$ \\
\enddata
\tablenotetext{1}{Column densities obtained from \citet{2}}
\end{deluxetable}

\section{Empirical Instrument Response}
\label{S:3}

As discussed in \hyperref[S:1.1]{Section 1.1}, the beam filter in the LETGS on \textit{Chandra} contains carbon. As such, it is imperative that we understand how this carbon interferes with the observed carbon K-edge of any observed object. To model the instrument response, we used a bright object with low column density so as to isolate absorption by the blocking filter. As per \hyperref[S:1.2]{Section 1.2}, this object is Mrk 421. 

\begin{figure}
    \centering
    \includegraphics[width=\linewidth]{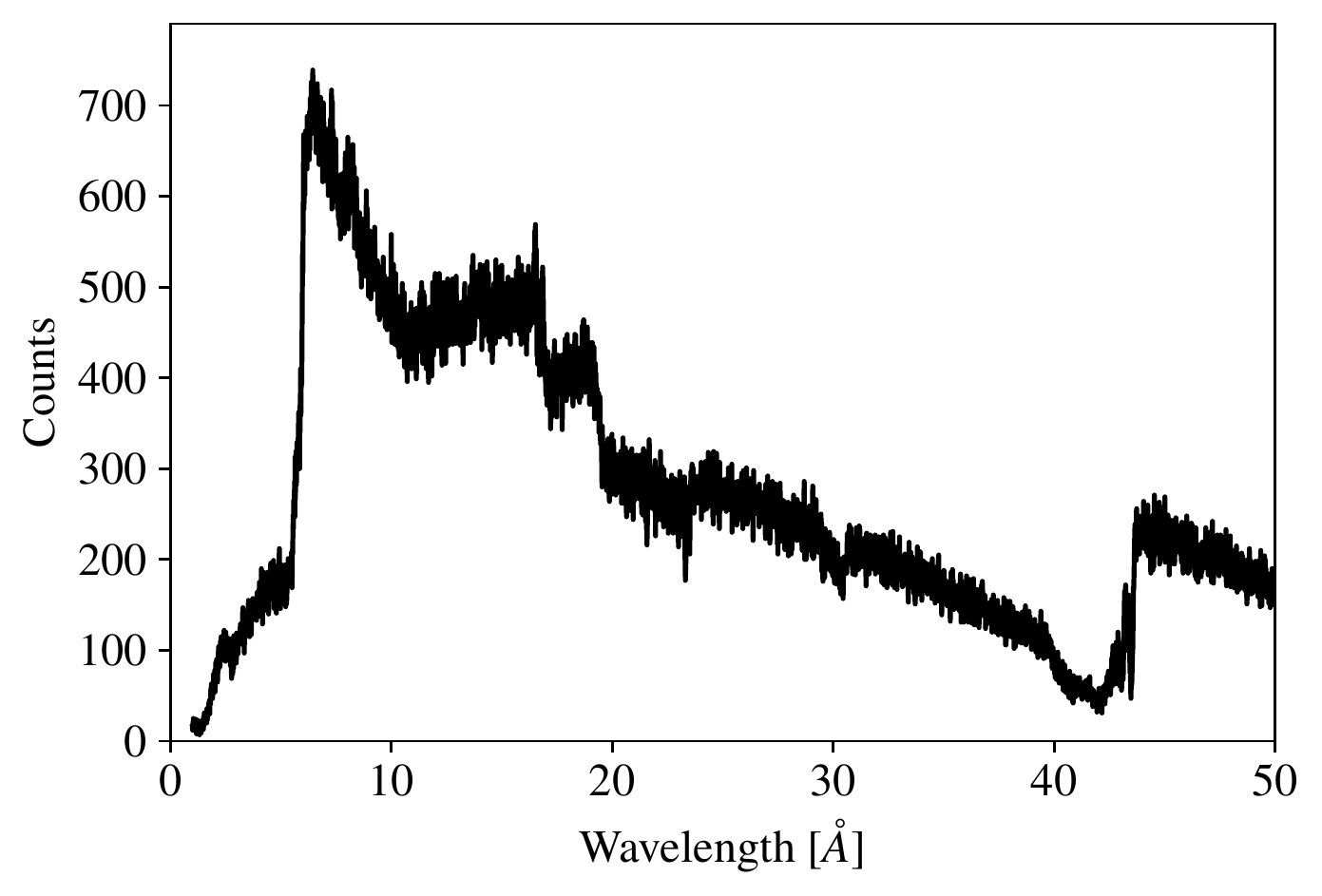}
    \caption{Positive Order Spectrum of Mrk 421. This shows the total counts of Mrk 421, in 0.05 $\mathring{\text{A}}$ bins. Note that the object is relatively bright around $15 \mathring{\text{A}}$, which means that the third order contribution around the C K edge at $45 \mathring{\text{A}}$ is not negligible.}
    \label{fig:3}
\end{figure}

The HRC camera does not have sufficient intrinsic energy resolution to separate the different grating diffraction orders. Instead, a quantitative analysis of the spectrum has to model the higher-order contribution at any given wavelength by modeling the broad-band spectrum and using the calibration of the LETG higher order efficiencies.
This analysis was done in CIAO as it afforded us the ability to examine multiple grating order spectra individually and in sum, at once, and perform a fit to determine which orders contributed the most \footnote{Fruscione, A., McDowell, J.C., Allen, G.E., et al. 2006, SPIE, 6270, 60}. The extracted broadband positive diffraction order spectrum of Mrk 421, binned to the approximate instrumental resolution of 0.05 $\mathring{\text{A}}$, is shown in \hyperref[fig:3]{Figure 2}. Note the fact that the object is relatively bright around $15 \mathring{\text{A}}$, which means that the third order contribution around the C K edge at $45 \mathring{\text{A}}$ is not negligible.

We briefly examine the contributions of the different spectral orders at wavelengths just short of the C K edge in some detail, to establish the relative importance of the higher-order contamination, and the influence of any uncertainties in that contribution on our estimate of the bare first-order flux. We compute the counts, $C_n(\lambda)$, in order $n$, observed at the position of first-order wavelength $\lambda$ in a bin of width $\Delta\lambda$, using the diffraction efficiency in the \textit{Chandra} Proposer's Observatory Guide. We assume that the first-order spectrum, $C_1(\lambda/n)$, at $\lambda/n$ is `clean' (no higher-order contribution from shorter wavelengths). For the case of Mrk 421 at $\sim 40 \mathring{\text{A}}$, this is a good assumption: the first significant higher-order contribution is third-order photons from 
$\sim 15 \mathring{\text{A}}$, and the spectrum at $15 \mathring{\text{A}}$ is not significantly contaminated by higher-order photons ({\it e.g.} third-order $5\mathring{\text{A}}$ photons) because the LETGS efficiency rapidly decreases at $\lambda < 5 \mathring{\text{A}}$.
Let the efficiency of the $n^{\text{th}}$ order be $e_n(\lambda)$ for wavelength $\lambda$. 
We get
\begin{equation}
    C_n(\lambda) = \frac{1}{n}C_1(\lambda/n)\frac{e_n(\lambda/n)}{e_1(\lambda/n)}
\end{equation}
where the factor $1/n$ accounts for the fact that the first order bin at $\lambda/n$ is stretched by a factor $n$ in order $n$. The reliability of our spectrum is measured by the number of contaminating counts versus total counts. We display the results for the positive orders in \hyperref[tab:2]{Table 2} for all objects of study. The counts are in a $0.05 \: \AA$ bin. Since the background is negligibly small in all objects, we clearly detect significant flux in first order at the wavelengths of the deepest minimum in the C K absorption at $42\ \mathring{\text{A}}$ in all objects. We compute the uncertainty by 
computing the Poissonian fluctuation in the counts at the wavelength $\lambda/n$ that is contaminating the first order.

\begin{deluxetable}{ccc}
\tablecaption{Orders $n=2-8$ Contributions at $42 \: \mathring{\text{A}}$ \label{tab:2}}
\tablehead{
\colhead{\hspace{0.7cm}Object} & \colhead{\hspace{0.7cm}Counts\tablenotemark{\small a} in $n=2-8$} & \colhead{\hspace{0.7cm}Total Counts}\\\vspace{-0.5cm}}
\decimals
\startdata
Mrk 421 & $28.22 \pm 3.84$ & 54.00\\
Ark 564 & $1.65 \pm 0.83$ & 4.00 \\
Mrk 509 & $1.42 \pm 0.82$ & 3.00 \\
1ES 1553 & $1.87 \pm 0.76$ & 6.00 \\
\enddata
\tablenotetext{\tiny a}{counts in a $0.05 \mathring{\text{A}}$ bin in positive orders}
\end{deluxetable}

\begin{figure*}
    \centering
    \includegraphics[scale=0.62]{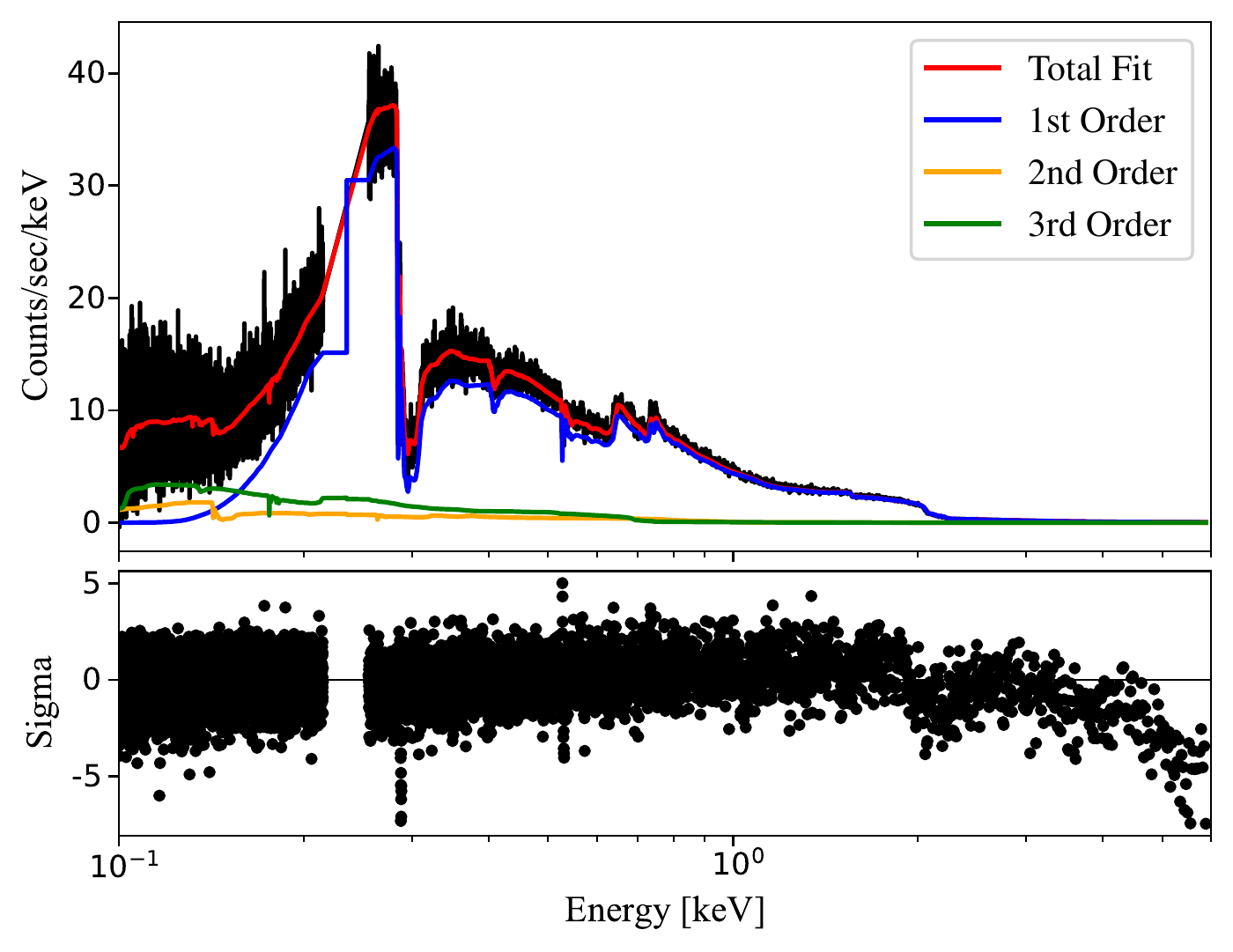}\hspace{0.25cm}\includegraphics[scale=0.62]{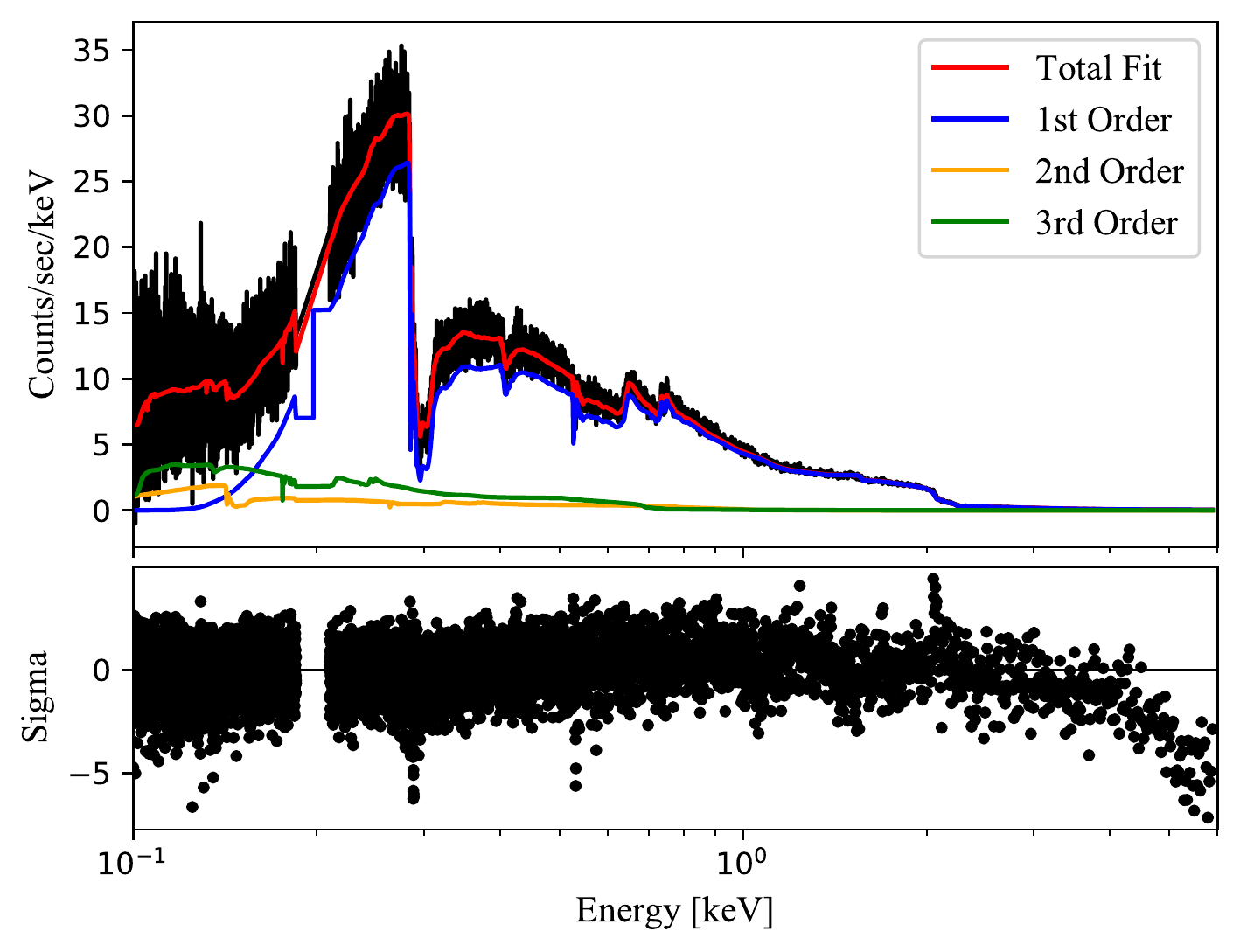}
    \caption{Higher Order Contributions of Mrk 421. We fit the positive (left) and negative (right) spectra of Mrk 421 with an absorbed power law, and demonstrate that there is a significant contribution from the third order, shown in green. The first order response is shown in blue, while the total fit is shown in red. This is then used to isolate the first order.}
    \label{fig:4}
\end{figure*}

A second subtlety arises from the difference between the positive and negative orders. To correctly analyze the total spectrum, we fit Mrk 421 positive and negative orders separately with an absorbed power law, which is then used to predict higher order contributions, using models from \verb!xspec! v12.9.1 \citep{1}. In \verb!xspec! terminology, this reads
    \verb!tbabs*zpowerlw!
where \verb!tbabs! models galactic absorption and \verb!zpowerlw! is a power law that accounts for the redshift. We set the galactic column density to $N_{\text{H}} = 1.92 \times 10^{20}$ cm$^{-2}$ and the redshft to $z = 0.031$ \citep{2, 3}. The fitting was done using CIAO, focusing on energies in the $0.1$ to $6$ keV range. The fit yielded a fit statistic $\chi^2/\text{d.o.f} = 1.354$, for 17558 degrees of freedom. This characterizes a reasonable fit for the present purpose, especially considering the simplicity of the model. The parameters for the corresponding fits in \hyperref[fig:4]{Figure 3} are contained in \hyperref[tab:3]{Table 3}. 

\begin{deluxetable}{ccc}
\tablecaption{Fit Parameters: Mrk 421 \label{tab:3}}
\tablehead{
\colhead{\hspace{0.7cm}Number}\hspace{0.7cm} & \colhead{\hspace{0.7cm}Parameter\tablenotemark{\small a}}\hspace{0.7cm} & \colhead{\hspace{0.7cm}Value}\hspace{0.7cm}\\\vspace{-0.5cm}}
\decimals
\startdata
Both\\
\hline
(1) & $N_{\text{H}}$ & $1.92 \times 10^{20}$ cm$^{-2}$ \\
(2) & $z$ & $0.031$ \\
\hline
Positive Order\\
\hline
(1) & $\Gamma$ & $2.093 \pm 0.003$ \\
(2) & $A_\Gamma$ & $0.405 \pm 0.0008$ A$_0$ \\
\hline
Negative Order\\
\hline
(1) & $\Gamma$ & $2.115 \pm 0.003$ \\
(2) & $A_\Gamma$ & $0.406 \pm 0.0008$ A$_0$ 
\enddata
\tablenotetext{\tiny a}{$N_{\text{H}}$ is the galactic column density and $z$ the redshift. $\Gamma$ is the photon index and $A_\Gamma$ is the normalization, which has units of A$_0$: photons keV$^{-1}$ cm$^{-2}$s$^{-1}$ at 1 keV.}\vspace{-0.1cm}
\end{deluxetable}

It is evident that the fit is good up until very high energies, when the zero order becomes relevant. Using these fit parameters, we modeled the higher-order contributions in the 0.25-0.35 keV band in order to isolate the first order response, in both positive and negative orders separately. 

In order to transform the first order spectrum into a proxy for the instrument response, we need to account for residual interstellar absorption. In the band of interest, the absorption is dominated by H, He, and C. In order to account for this absorption, we used \citet{29} values for the photoelectric absorption cross section. The optical depth is $\tau = N_{\text{H}} \sigma(E)$ where $\sigma(E)$ is the absorption cross section per H atom and $N_{\text{H}}$ is the column density of H atoms. We then divide by $e^{-\tau}$. The optical depth due to photoelectric absorption by the neutral ISM varies smoothly by a factor $\sim 2.5$ across the 0.25-0.35 keV band; the transmission rises from $\sim 0.4$ to $\sim 0.7$. The optical depth due to residual interstellar C K absorption, assuming normal abundances, is $\approx 0.05$, at energies just above the absorption edge. Absorption from all other elements is negligibly small at these energies \citep{29}.

\begin{figure}
    \centering
    \includegraphics[width=\linewidth]{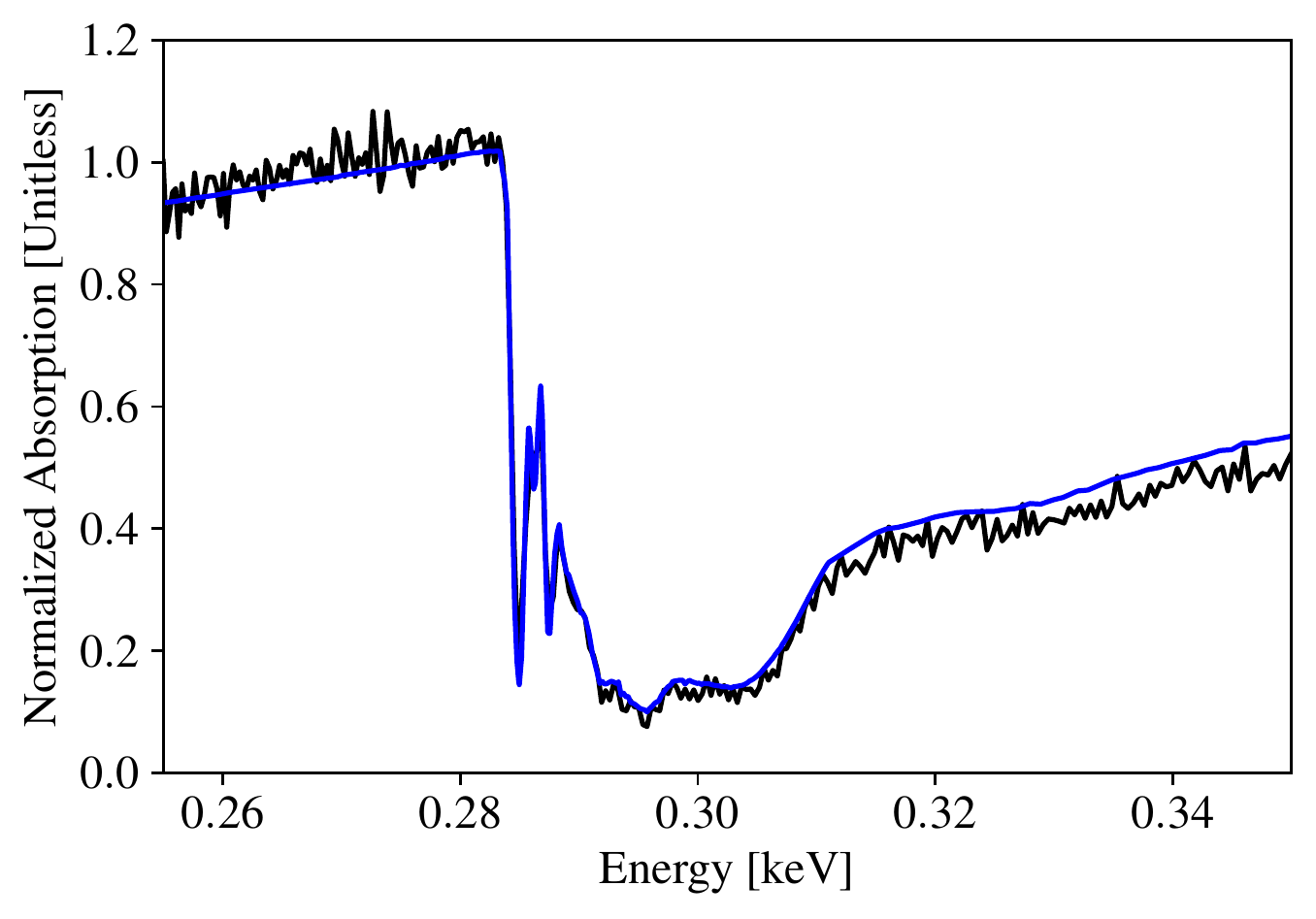}
    \caption{Empirical Instrument Response, shown in black, normalized to 1 at 0.27 keV of the LETGS in the region around the carbon K edge, derived from the spectrum of Mkn 421. The blue line corresponds to synchrotron measurements of the instrumental features of the UVIS, demonstrating good agreement in the 280-300 eV energy range of interest.}
    \label{fig:5}
\end{figure}

Lastly, we divide the positive and negative orders by their power law slope $E^{-\Gamma}$. Doing so for each, up to a normalization factor, will isolate the contributions from instrumental carbon absorption alone. The final model histogram for the total instrument response is shown in \hyperref[fig:5]{Figure 4}, where we normalized the response to $1$ at $0.27$ keV. In all subsequent data analysis, we will use this model to account for absorption by the instrument and refer to it as the \textit{Empirical Instrument Response Function} (EIRF). In that same figure, we overlay synchrotron measurements of the transmission of the UVIS shield\footnote{For reference: http://hea-www.harvard.edu/HRC/calib/uvismodel.html}, demonstrating good agreement. We note that, as per \citet{4}, there is extra C II-IV absorption in the spectrum of Mkn 421 not properly accounted for in our models. 

\section{Interstellar Absorption Spectra}
\label{S:4}
Now that we have an EIRF, we are in a position to determine whether there is significant absorption in either Ark 564, Mrk 509, or 1ES 1553. We will find that there are non-trivial absorption features in the spectrum for 1ES 1553. We begin by isolating the first order contribution to each of the spectra. This is done by fitting an absorbed power law to the positive and negative orders separately. 

Next, we take the first order spectrum and isolate the absorption due to carbon. This is done in the same way as Mrk 421, by removing absorption due to hydrogen and helium as well as the intrinsic powerlaw slope. What remains is carbon absorption either due to the instrument or the ISM. A comparison of the EIRF to each of the interstellar absorption spectra is shown in
\hyperref[fig:6]{Figure 5} and \hyperref[fig:7]{Figure 6}. It is evident that the 1ES 1553 interstellar absorption spectrum has the most significant difference with the EIRF. To the low energy side of the main $n = 1 - 2$ absorption line at 0.285 keV, we see a gradual `shoulder' in the transmission spectrum. We will attribute this residual difference to interstellar absorption (extinction, really)  and refer to it as the \textit{red shoulder}.

\section{Analysis of the Interstellar Absorption Spectrum of 1ES 1553}
\label{S:5}

\begin{figure}
    \centering
    \includegraphics[width=\linewidth]{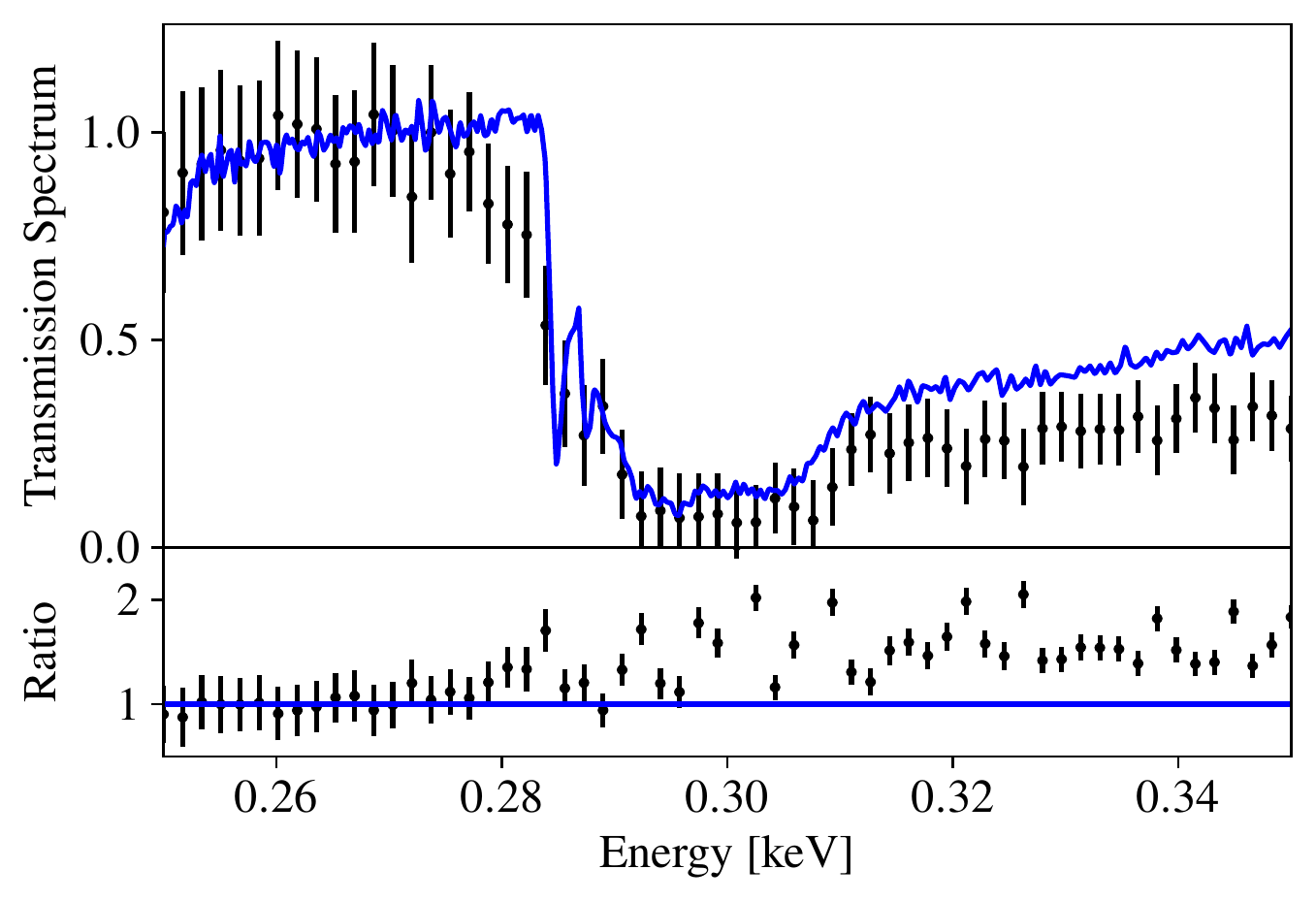}
    \caption{Residual absorption in the C K band in 1ES 1553+113 observed with LETGS. In the top panel, we show the empirical transmission spectrum along the line of sight to 1ES 1553 in black (positive and negative spectral orders averaged). If the absorption was solely due to the instrument, it would match the empirical instrument response function, shown in blue, but there is a different signature, which we call the ``red shoulder." The ratio of the spectrum of 1ES 1553 to the empirical response function is shown in the bottom panel.}
    \label{fig:6}
\end{figure}

With our new empirical instrument response model, we are ready to study a source with significant interstellar absorption, 1ES 1553+113. 
The spectrum of this source, after our reduction procedure described above, if there was no interstellar carbon absorption along the line of sight to 1ES 1553, would be the same as the empirical instrument response we found for Mkn 421. However, we find a noticeable difference, as shown in \hyperref[fig:6]{Figure 5}. To the low energy side of the main $n=1-2$ absorption line at 0.285 keV, we see the aforementioned red shoulder in the transmission spectrum. 

We first discuss possible sources of systematic error that could have produced the apparent absorption on the red side of the $n=1-2$ transition. We visually inspected the HRC-S spectral images around $43-45$ \AA\ ($\sim 275-290$ eV) in both positive and negative spectral orders, but saw no anomalies. As far as we are aware, there are also no known areas of uncalibrated, significantly reduced quantum efficiency of the HRC-S in these regions. The positive and negative spectral orders both show the red shoulder in 1ES 1553, and are statistically consistent with each other. We also verified that there has been no secular change or degradation in the UVIS, or in fact in any other instrument component. This was done by comparing spectra of Mrk 421 obtained both prior to the observation of 1ES 1553, and after the observation of 1ES 1553. The observation IDs for the latter observations, obtained through publicly available \textit{Chandra} database, are 14322, 14323, and 14324. Next, there is a possibility that an observation using a non-default dither pattern might have caused the relevant parts of the spectrum to be projected onto a region of the UVIS 
with properties different from those in the Mrk 421 observations (see the schematic of the 
HRC layout at https://cxc.cfa.harvard.edu/proposer/POG/html/ chap7.html\#tth\_sEc7.2). However, for 1ES-1553 all data was taken at the same, default, position and, taking the telescope dithering into account, all photons passed through the same portion of the UVIS shield, with the same 2750\AA\ of polyimide (the section marked 'S2' in the Figure in the Proposers Observatory Guide quoted above). In addition, while the aluminum layer in the 
UVIS has varying thicknesses as a function of position, aluminum has no spectral features in the 250-350 eV band.

\begin{figure}
    \centering
    \includegraphics[width=\linewidth]{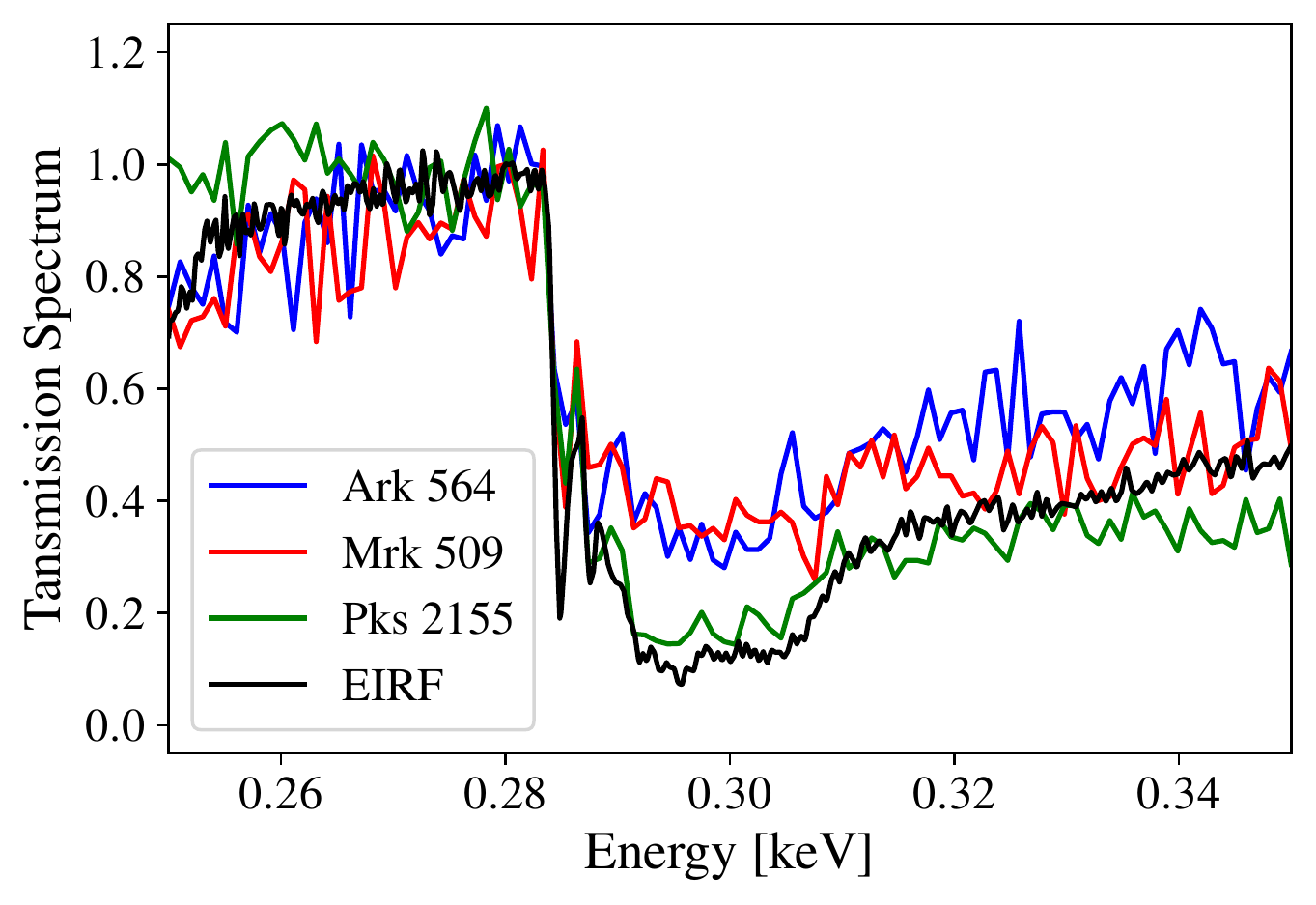}
    \caption{Comparison of EIRF to other objects. Shown in the figure are the transmission spectra of Ark 564, Mrk 509, and PKS 2155 all normalized at $0.28$ keV. Superimposed on the spectra is a similarly normalized empirical instrument response function shown in black. There is no visible 'red shoulder' as seen for 1ES 1553. Note that the depth of the continuum C K absorption in Ark 564 and Mrk 509 appears shallower than the instrumental absorption; this probably signals that there is residual higher spectral order contamination in these spectra in the $285-340$ eV range.}
    \label{fig:7}
\end{figure}

Finally, and most directly, a red shoulder is not seen in any other continuum spectrum of sufficient statistical quality that we have inspected. In other words, all other objects of study (Mrk 509, Ark 564, and PKS 2155) display features that are very similar to Mrk 421. This is shown in \hyperref[fig:7]{Figure 6}. This rules out an instrumental origin for the observed `red shoulder.' Note that the depth of the edge is not the same across different objects due to relatively higher signal to noise and differences in power laws for the fit. This, however, does not affect the presence of a red shoulder feature.

Next, we considered the possibility of accidental superposition of an intrinsic absorption feature in 1ES 1553 or the intervening intergalactic medium. For an assumed redshift $z > 0.395$ for 1ES 1553 \citep{40}, the rest frame energy $E_0$ of a feature at 280 eV is 
$E_0 > 391$ eV. That excludes the strongest expected carbon lines, \ion{C}{6} Ly$\alpha$ ($E_0 = 368$ eV), and the \ion{C}{5} $n=1-2$ resonance line ($E_0 = 308$ eV). The next expected strong feature would be the \ion{O}{7} $n=1-2$ resonance line ($E_0 = 574$ eV), which would imply a source redshift of $z=1.05$. That would place the corresponding \ion{O}{7} $ n=1-3$ resonance line at 325 eV and \ion{O}{8} Ly$\alpha$ at 319 eV; no significant absorption lines are seen at these energies in the spectrum of the source.
Likewise, an intergalactic absorption feature from the K-shell ions of C and O of the equivalent width we see in \hyperref[fig:6]{Figure 5} appears unlikely. The observed equivalent width of the red shoulder (EW $\approx 3-4 $ eV or $\approx 500-600$ m\AA) is nearly two orders of magnitude larger than that of the intergalactic \ion{O}{7} lines that have been identified to date \citep{51}. A very heavily saturated and/or broadened $n=1-2$ line would surely also have produced detectable absorption in higher series members. 

If we therefore proceed on the assumption that the red shoulder is due to extinction in the interstellar medium along the line of sight to 1ES 1553, it is natural to suspect a form of carbon to be the source of the extinction. The shape of the feature, a fairly wide `dip' just to the long-wavelength side of the $1s-2\pi^*$ resonance in neutral carbon in graphene or graphite at 285 eV in fact directly mimics the behavior of the real part of the index of refraction, responsible for scattering, of a simple harmonic oscillator (the response in the imaginary part of the index, responsible for true absorption, is much narrower) \citep[pg. 311]{44}. This implies that we are seeing the effect of scattering by large-sized particles, large enough to cause a finite phase shift of the radiation on passage through the particles. The photoabsorption by PHAs, while it does exhibit finite broadening and additional structure, on the 1s-2$\pi$* transition and at slightly higher energies \citep{Reitsma, Klues} cannot match the width of the red shoulder. In addition, the PHAs are too small to cause much scattering.

If that interpretation is correct, we are seeing the effect of extinction by scattering of radiation, not absorption. For particle sizes $a < 15\mu$m, characteristic scattering angles for X-rays of wavelength $\lambda = 44$ \AA\ are of order $\lambda/a > 1$ arcminutes. Scattering will therefore effectively remove radiation from the beam, placing most of it far outside the {\it Chandra} PSF, and therefore far away from the dispersed light image on the HRC-S camera. The effect will be to cause a dip in the transmission spectrum of the interstellar medium, when considering light concentrated in small apertures of order a few arcsec in diameter. 
As a very rough estimate for the apparent efficiency of this extinction, we note that a circular aperture with a diameter 20 arcsec, of order the width of our rectangular spectral extraction region on the HRC-S, contains only a fraction $10^{-5}$ of the intensity of 
a scattering halo produced by 280 eV photons on grains of radius $0.1$ $\mu$m. 

\citet{5} shows calculations for the index of refraction of graphite spheres that display exactly the behavior of the real part of the index of refraction described above, in his Figure 3. We therefore proceed on the assumption that we are detecting the effect of scattering by graphite grains in the spectrum of 1ES 1553.

\begin{figure}
    \centering
    \includegraphics[width=\linewidth]{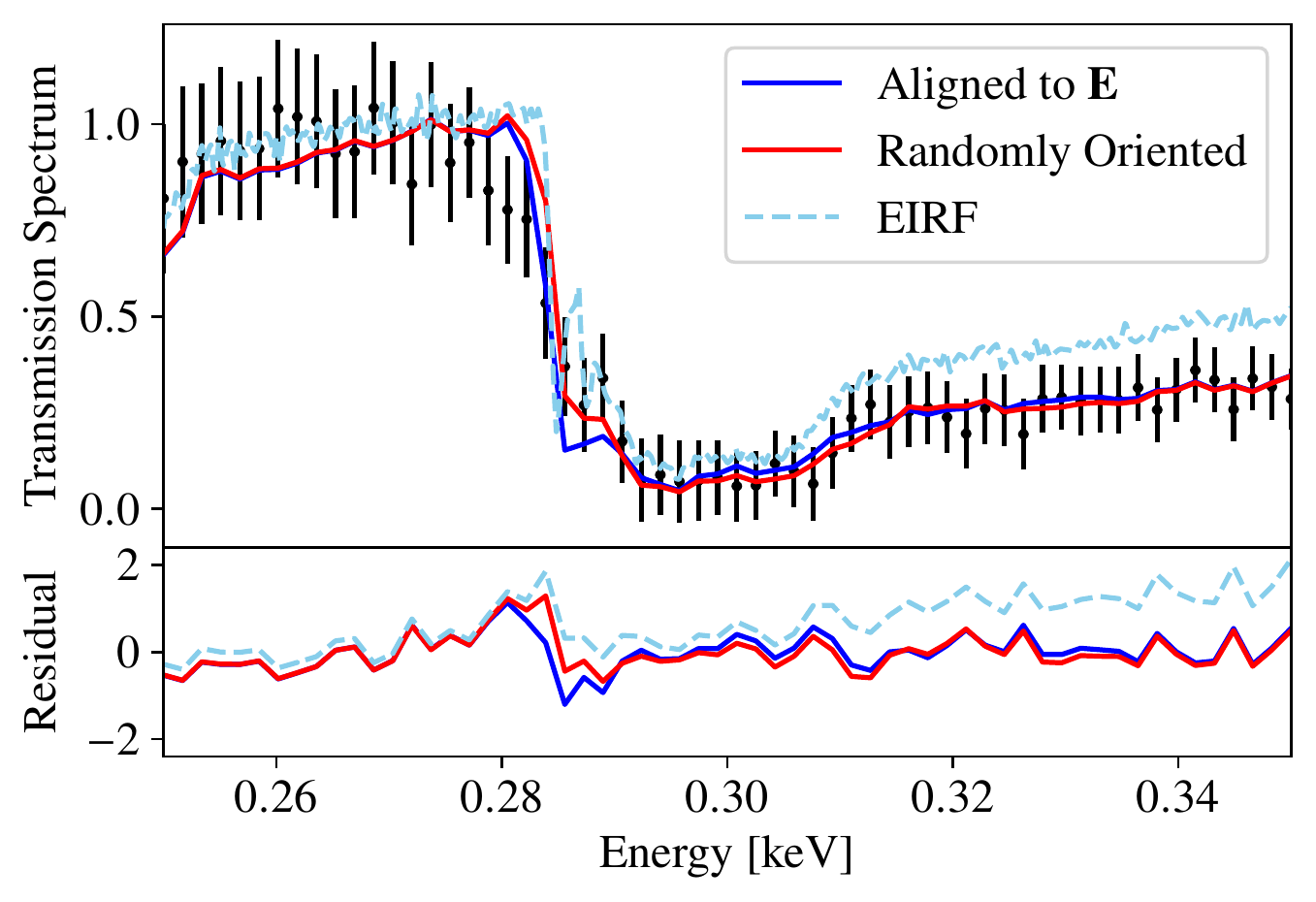}
    \caption{Fit of the ISM carbon K transmission to the spectrum of 1ES 1553. We use both the instrument response model as well as the transmission spectrum for the extinction of light from Mie scattering on graphite particles. This fit is good and demonstrates that the red shoulder can be explained by graphite particles of radius $\sim 0.1$ $\mu$m and a carbon column of $4 \times 10^{8}$ cm$^{-2}$. The blue line corresponds to a set of particles in which the graphene planes are perpendicular to the electric field, while the red line employs a set of randomly oriented grains using the '1/3-2/3' approximation. The red line is significantly offset from the data, by 2-3 eV to higher energies. The dotted light blue line shows that contribution from the EIRF to both fits. The measured spectrum favors the extinction by graphene planes aligned to polarized radiation (the blue line). The lower panel shows the residual, being the observed data subtracted from the model divided by the error for each model.
    }
    \label{fig:8}
\end{figure}

Graphite is a strongly anisotropic material. The scattering cross section will be largest for X-ray radiation for which the electric vector is parallel to the vector normal to the graphene planes that make up the graphite (``{\bf E}$//$ {\bf \^c}").
From the optical constants given by \citet{5}, we compute the total transmission of a fixed mass column density of graphite dust particles. For exploratory purposes, we assume spherical grains, of a single radius $a$. We fix the total carbon mass column density (g cm$^{-2}$) at $N_{\rm C}m_{\rm C} = A_{\rm C} m_{\rm C} N_{\rm H}$, with $N_{\rm H} = 3.65 \times 10^{20}$ cm$^{-2}$ the (number) column density of H atoms towards 1ES 1553, $A_{\rm C} = 2.9 \times 10^{-4}$ the abundance of carbon by number \citep{45}, $N_{\rm C}$ the carbon number column density, and $m_{\rm C}$ the mass of a carbon atom. We have $N_{\rm C} m_{\rm C} = 2.1 \times 10^{-6}$ g cm$^{-2}$. The number column density of graphite grains is $N_{\rm grains} = N_{\rm C} m_{\rm C}/((4\pi/3)a^3 \rho_{\rm graphite}) = 2.2 \times 10^{8}$ cm$^{-2}$, where we have used $\rho_{\rm graphite} = 2.26$ g cm$^{-3}$ and $a = 0.1$ $\mu$m. 
We calculate the scattering and absorption cross sections using a Mie code \citep{46} under the assumption that the X-rays are linearly polarized, and the graphite grains are aligned ($\mathbf{E}//\mathbf{\hat{c}}$). We display the resulting transmission spectrum in blue with a fitted grain column density of $N_{\rm grains} = \left(4.0 \pm 0.5\right) \times 10^8$ cm$^{-2}$ for grains of radius $a = 0.1\mu$m in \hyperref[fig:8]{Figure 7}. The transmission spectrum fits the data well, correctly accounting for the presence of a red shoulder. Interestingly, the extinction by scattering also reproduces the observed 290-350 eV spectrum (in the C K continuum) well. 

Note that the column density we used ($N_{\rm grains} = 4 \times 10^8$ cm$^{-2}$) implies a carbon mass column density $N_{\rm C} m_{\rm C} = 8.4 \times 10^{-6}$ g cm$^{-2}$, about four times the C mass column density we estimated for the measured H column density of $N_{\rm H} = 3.65 \times 10^{20}$ cm$^{-2}$. If we allow for unpolarized radiation and randomly oriented grains (or rather, graphene planes) the effective scattering cross sections are lower than than what we assumed above. \citet{5} discusses approximate calculations for randomly oriented graphite grains (the '1/3-2/3 approximation'). Directly comparing the ``{\bf E}$//$ {\bf \^c}" cross sections and the cross sections for randomly oriented grains, we find that the latter are a factor 2-3 smaller. If we take the grain radius $a=0.1\mu$m to be representative of grains with radii in the range $a=0.1-0.2\mu$m, we find that for a 'normal' distribution of grain sizes, only about $1/4$ of the total mass in grains has radii in that range (assuming a Mathis-Rumpl-Nordsieck distribution of grain radii, between $a=10^{-3}\mu$m and $a=0.3\mu$m; \citet{50}). Finally, not all carbon will be in grains. \citet{49} find dust mass fractions for C along 21 Galactic sightlines in the range $\approx 0.2-0.85$, so nearly all C could be in grains along our line of sight, but in the more typical case (dust mass fraction for C $\approx 0.6$), there would be an extra $\approx 70\%$ in carbon gas. Taking all these factors together, we find that our column of $N_{\rm grains} = 4 \times 10^8$ cm$^{-2}$ of grains of radius $a\sim 0.1\mu$m implies a total C mass column density of $> 4 \times 4 \times (1/0.85) \times 4.2 \times 10^{-6} = 4.0 \times 10^{-5}$ g cm$^{-2}$, a factor $\gtrsim 40$ larger than we expect for a 'normal' ISM sightline of H column density $N_{\rm H} = 3.65 \times 10^{20}$ cm$^{-2}$. In summary, we find that the observed amplitude of the red shoulder can be explained by the scattering of X-rays by a column of $N_{\rm grains} = 4 \times 10^8$ cm$^{-2}$ of $0.1-0.2\mu$m grains, if all the grains are aligned {\bf E}$//$ {\bf \^c}, and the X-rays are polarized. If the sightline had only large grains ($a = 0.1-0.2\mu$m) this might just be possible with the mass of carbon estimated to be present along this line of sight (to within a factor of a few, which is probably not excluded by variations in the carbon abundance in the ISM).

In addition to the above, we also compare the {\it shape} of the transmission curve in the 1/3-2/3 approximation for the cross sections for randomly oriented grains, to the data, in \autoref{fig:8}. We obtain an approximate fit to the data with grains of size $a = 0.1$ $\mu$m and a fitted column $N_{\rm grains} = \left(3.9 \pm 0.5\right) \times 10^8$ cm$^{-2}$.
However, this transmission curve has a significant offset in energy with respect to the measured spectrum: the steep decline in transmission on the red side of the $1s-2\pi^*$ is offset to higher energies with respect to the data by $\sim 2-3$ eV, due to the fact that the pure scattering part of the extinction is relatively less important than for aligned grains. This offset is much larger than the accuracy of the wavelength scale of the LETGS (which is a small fraction of the instrument resolution of $\Delta\lambda = 0.05$ \AA\ or $\Delta E = 0.35$ eV at 300 eV), or any conceivable radial velocity Doppler offset. 
In fact, the transmission for graphene planes {\it aligned with polarized X-rays} actually fits the data better.

\begin{figure}
    \centering
    \includegraphics[width=\linewidth]{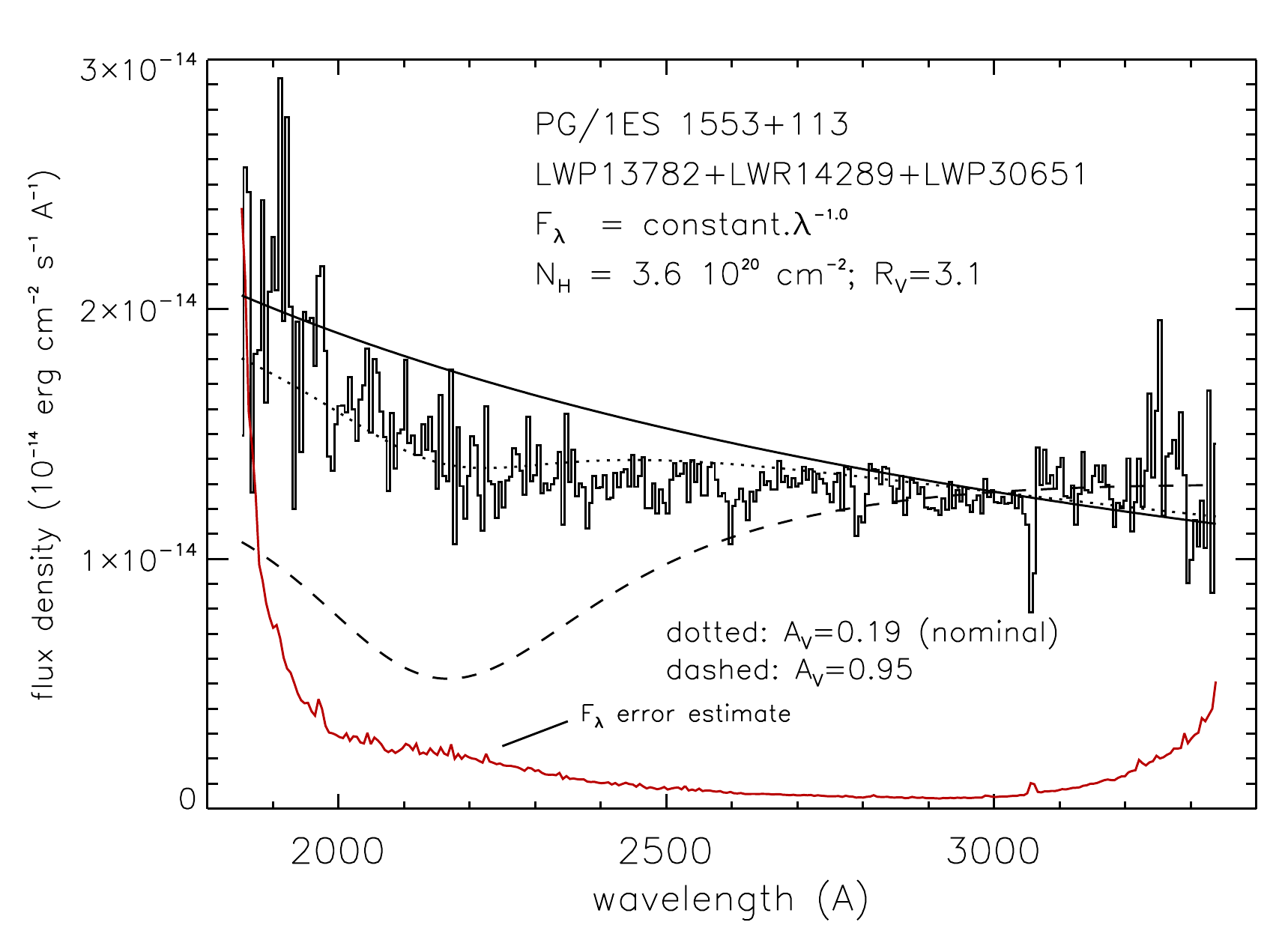}
    \caption{Near-UV spectrum of 1ES 1553. Shown is the combined spectrum from three observations listed in \hyperref[tab:4]{Table 4}. The size of the error bars is shown by the red line. The solid line shows a smooth powerlaw ($F_{\lambda} \propto \lambda^{-1.0}$). The dotted line shows this powerlaw with extinction by a nominal Galactic ISM column density of $N_{\rm H} = 3.65 \times 10^{20}$ cm$^{-2}$, cosmic carbon abundance, and $R_V = 3.1$; the optical extinction for this column is $A_V = 0.19$. The dashed line shows the same, but for a five times larger column density. All models have been normalized at 3000 \AA.
    }
    \label{fig:UVfig}
\end{figure}

In an attempt to compare to other signatures of interstellar carbon, we inspected the ultraviolet spectrum of 1ES 1553. Around 2175 \AA, a 'bump' in the extinction is thought to arise from extinction by very small carbon-rich particles, $a \sim 10^{-3}-10^{-2}\mu$m. This is likely to be a population of very large carbon-based molecules, usually referred to as Polycyclic Aromatic Hydrocarbons (PAHs). With the simultaneous detection of the UV extinction bump and the detection of the red shoulder in the C K absorption spectrum, we have the means to independently constrain the small and the large ends of the grain size distribution (large grains do not contribute much to the UV extinction; small grains do not contribute much to the red shoulder).
1ES 1553 was observed with the {\it International Ultraviolet Explorer}. We extracted processed spectra from the archive\footnote{{\tt https://archive.stsci.edu/iue/}}. Three exposures cover the $1800-3000$ \AA\ range with significant signal-to-noise; these are listed in Table 4. 
We added the background-subtracted and fluxed spectra, weighting them with the inverse square of the estimated error, and binned the spectrum in 5.33 \AA\ bins. The result is shown in 
\hyperref[fig:UVfig]{Figure 8}. The data are noisy at the extremes of the wavelength band, but appear to be roughly consistent with a flat or moderately rising spectrum ($F_{\lambda} \propto \lambda^{\alpha}$, $\alpha = -1.0\ \textendash\ 0.0$). We superimposed extinction using the nominal extinction curve given by \citet{48}, computed for a column density of $N_{\rm H} = 3.65 \times 10^{20}$ cm$^{-2}$, $N_{\rm H}/E(B-V) = 5.8 \times 10^{21}$ H cm$^{-2}$ mag$^{-1}$, and $R_V = A_V/E(B-V) = 3.1$. The model was arbitrarily normalized at $3000$ \AA. This continuum model is roughly consistent with the measured spectrum, indicating that the UV extinction along this line of sight is not very different from the average interstellar UV extinction in the Galactic disk. We also show the continuum with extinction as would be present if the carbon column was five times larger than the nominal value. The resulting prominent `2175 \AA\ bump' is not present in the measured spectrum. We conclude that if the large grains that are responsible for the observed scattering in the X-ray spectrum were part of a `normal' ISM grain population, there should be a pronounced 2175 \AA\ feature in the UV spectrum.

\begin{deluxetable}{cccc}
\tablecaption{UV Observations\label{tab:4}}
\tablehead{
\colhead{\hspace{0.34cm} ObsID}\hspace{0.34cm} & \colhead{\hspace{0.34cm} Galaxy}\hspace{0.34cm} & \colhead{\hspace{0.34cm} Date}\hspace{0.34cm} & \colhead{\hspace{0.34cm} Net exp. (ks)}\hspace{0.34cm}\\\vspace{-0.5cm}}
\decimals
\startdata
LWP13782 & 1ES+1553 & 08-02-1988 & 9600 \\
LWR14289 & 1ES+1553 & 09-29-1982 & 14400 \\
LWP30651 & 1ES+1553 & 05-09-1995 & 10200 \\
\enddata
\end{deluxetable}


\section{Summary and Discussion}
\label{S:6}

We have presented X-ray spectroscopic evidence that a significant fraction of interstellar carbon along the line of sight towards the blazar 1ES 1553+113 may be in graphite particles. After accounting for the absorption by neutral carbon in the instrument, using a model independent calibration based on the spectrum of a continuum source nearly free of interstellar C absorption, we find a broad extinction feature to the red side of the $1s-2\pi^*$ transition in neutral carbon. The functional form and amplitude of the feature are consistent with scattering by graphite spheres of radii 0.1-0.15 $\mu$m. We can discount explanations for the feature based on instrumental effects, or absorption intrinsic to the blazar, or absorption by the intergalactic medium. 

The interaction cross section of graphite is strongly anisotropic near the K-shell resonances.
We find that using the cross sections for graphite particles with orientation of the graphene planes perpendicular to the electric field quantitatively fits the observed spectrum very well. However, the carbon mass column density required to explain the amount of extinction already moderately (factor two) exceeds the available mass column density estimated from the measured neutral hydrogen column density, assuming nominal carbon abundance. If we allow for random orientation of the graphene planes and unpolarized X-rays, the discrepancy increases by another factor 2-3. Assuming the $a \sim 0.1-0.2 \mu$m grains to be part of a 'normal' population of interstellar grains with a wide distribution of grain sizes increases the discrepancy by a factor of several (a factor 4 for the canonical MRN distribution). 
Any finite depletion of carbon into dust short of full depletion makes the discrepancy wider yet.
We verified that there is no evidence in the UV extinction characteristics of the source for an anomalously large column density of very small interstellar carbonaceous grains (PAHs); this latter fact does not preclude the presence of a large column of large-sized grains.

There are six extragalactic continuum sources with sufficient signal to exhibit a nontrivial spectrum with LETGS in the {\it Chandra} archive. Five, including 1ES1553+113, are listed in \hyperref[tab:1]{Table 1}. Extensive exposure for a sixth, H1821+643, entered the public domain recently; a visual inspection of the spectrum around the C K edge for H1821+643 using {\tt tgcat}\footnote{{\tt tgcat.mit.edu}} shows its C K spectrum to be consistent with simply the shape of the EIRF, thus resembling the shapes for the sources shown in \autoref{fig:7} with the exception of 1ES1553+113. Since the C K spectrum of 1ES1553+113 appears to show a unique signature of interaction of the X-rays with the Galactic ISM, it makes sense to see if its line of sight has unusual properties with respect to its estimated dust column density and Galactic magnetic field strength. 

\begin{figure*}
    \centering
    \includegraphics[scale=0.325]{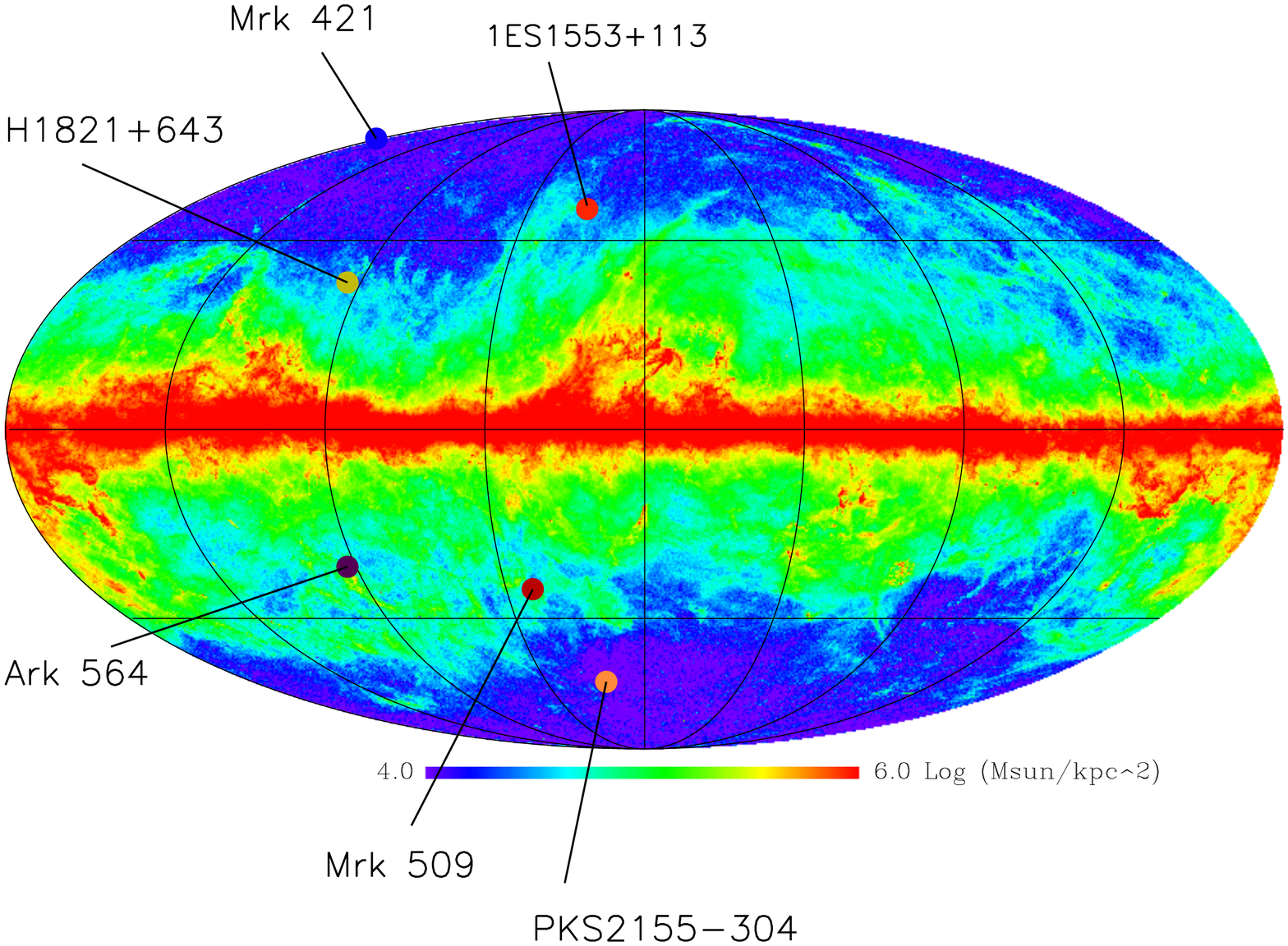}\includegraphics[scale=0.325]{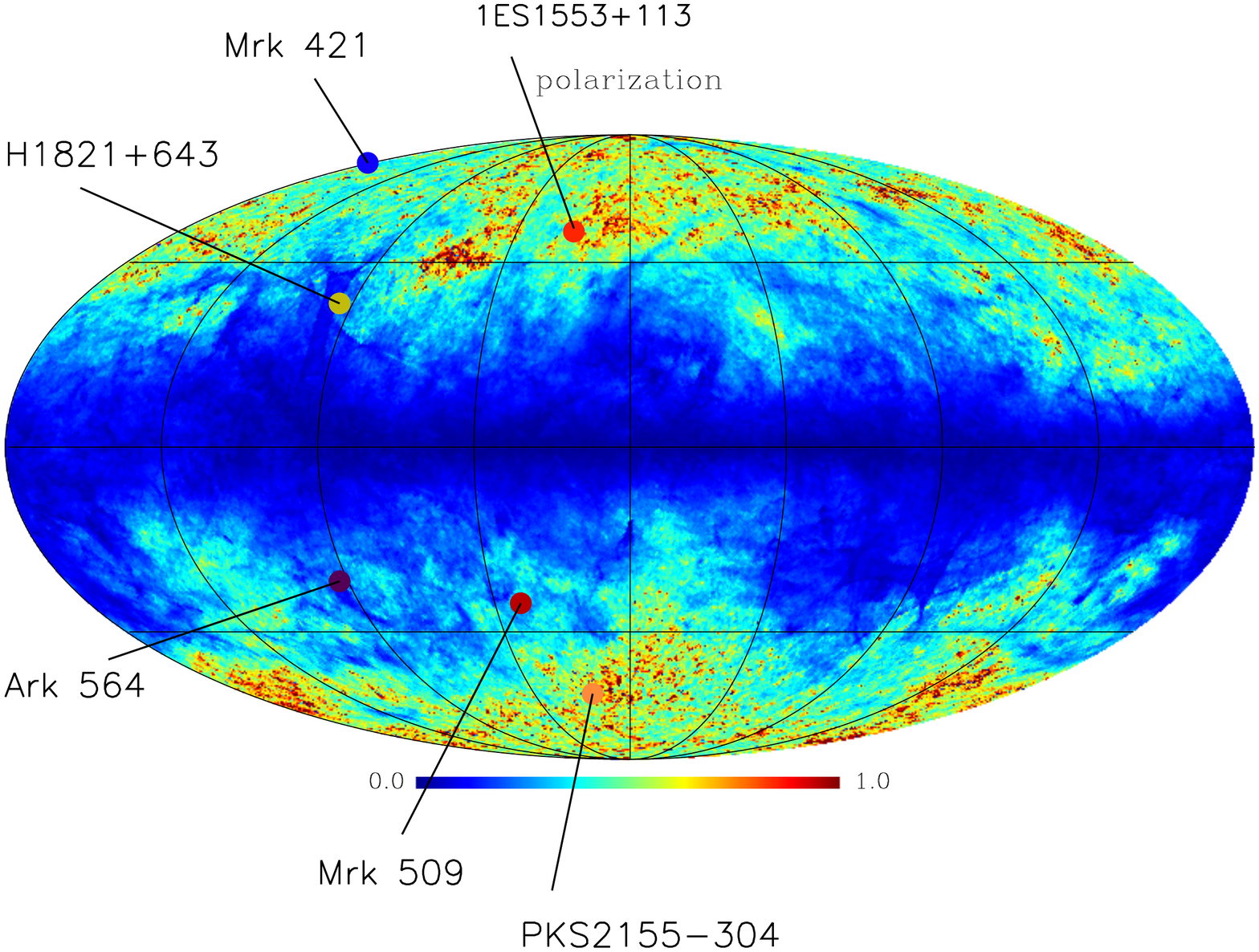}
    \caption{Sky Maps of the dust column and polarization. On the left panel, we display the Distribution of the Galactic dust column density, based on the Draine-Li broad-band emission model applied to combined {\it IRAS}, {\it WISE}, and {\it Planck} data (Ade et al. 2016). The scale is logarithmic, between $10^4-10^6 M_{\odot}$ kpc$^{-2}$, to focus on the intermediate range of column densities relevant to the positions of the six sources with a nontrivial {\it Chandra} LETGS C K absorption spectrum. In addition, polarized grains of dust can contribute to higher amounts of extinction. On the right panel, we show polarization measured as $p = \sqrt{Q^2 + U^2}/I$, written in terms of the Stokes parameters. There are high polarization values at higher galactic latitudes.}
    \label{fig:SkyMaps}
\end{figure*}

\autoref{fig:SkyMaps}a shows the column density of dust on the sky, calculated from a fit to dust emissivity (Draine and Li 2007) to the combined {\it IRAS}, {\it WISE}, and {\it Planck} data \citep{54}. As already suggested by the column densities of neutral H towards these six sources, the dust column densities are all in the range $10^5 M_{\odot}$ kpc$^{-1}$, with the exception of Mrk 421, which has a very low dust column. 1ES1553+113, however, is located near Loop I, a large coherent structure of enhanced density and magnetic field strength, outlined by emission from low-frequency radio synchrotron radiation to soft X-rays \citep[their Figure 21]{54}. In \autoref{fig:SkyMaps}b, we show the degree of polarization of 353 GHz mission observed with {\it Planck}, as a proxy for the strength of the component of the magnetic field projected on the plane of the sky. Loop I is visible again. A detailed magnetic field map near the position of 1ES1553+113 is shown by \citet[their Figure 25]{54}. While the evidence from dust emissivity and radio polarization is too sparse to draw any firm conclusions to differentiate 1ES1553+113 from the other five objects, it is not inconsistent with the idea that the line of sight to 1ES1553+113 might exhibit a relatively large polarization.

We conclude that the line of sight towards 1ES 1553 either has an excess column density of large-sized graphite particles (with a column mass density in carbon at least equal to the total carbon column expected in this line of sight), and/or shows the effect of scattering of polarized X-rays from 1ES 1553 by graphite particles with global alignment of the graphene planes. We note that the X-ray continuum emission from blazars is expected to be polarized, possibly as highly as 70 \% \citep{47}, though 1ES 1553 may not be among the sources with the highest expected polarization \citep{52}. Inspection of maps of the geometry of the Galactic magnetic field shows global structure in the direction towards 1ES 1553 (Galactic coordinates 
($l,b$) = (21.9, 44.0) degrees), but we are reluctant to speculate on an interstellar grain alignment mechanism that operates at the level of the graphite crystal planes.

With the current instrumentation, a full survey of the sky using all suitable extragalactic X-ray sources, other than the handful of very bright ones, is clearly impossible. One possible way to expand on the sample of six bright sources is to assemble a larger sample of fainter sources and investigate the C K absorption in a statistical sense, if possible differentially, by characteristics of the sight lines derived from other observations. In addition, the capabilities of the proposed Arcus mission, a sensitive high-resolution $0.2-1.2$ keV X-ray spectroscopy mission, are almost ideal for C K spectroscopy of the ISM (http://www.arcusxray.org/overview.html; \citet{36}).

\section*{Acknowledgements}

We thank David Huenemoerder for helpful advice on generating the correct spectral files for data analysis used throughout the paper. We would like to express our gratitude to our anonymous referee, whose extensive comments were very helpful in improving this paper. This research has made use of data obtained through the High Energy Astrophysics Science Archive Research Center Online Service, provided by the NASA/Goddard Space Flight Center.

\bibliographystyle{aasjournal.bst}
\bibliography{Bibliography.bib}

\end{document}